\documentclass[prl,twocolumn,letterpaper, superscriptaddress, longbibliography ]{revtex4-1}

\usepackage{graphicx}
\usepackage{dcolumn}
\usepackage{bm}

\usepackage{graphicx}
\usepackage{bm}
\usepackage{amssymb}
\usepackage{amsmath}
\usepackage{hyperref}
\usepackage{dcolumn}
\usepackage{subfigure}
\usepackage{threeparttable}
\usepackage{multirow}
\usepackage[usenames,dvipsnames]{color}
\usepackage{wasysym}
\usepackage{txfonts}
\usepackage{mathtools}

\usepackage{soul}
\usepackage{xcolor}
\usepackage[normalem]{ulem}
\newcolumntype{P}[1]{>{\centering\arraybackslash}p{#1}}

\usepackage{graphicx}
\usepackage{dcolumn}
\usepackage{bm}

\begin{document}

\preprint{APS/123-QED}

\title[title]{Even-Odd-Layer-Dependent Symmetry Breaking in Synthetic Antiferromagnets}

\author{M. M. Subedi}
\affiliation{Department of Physics and Astronomy, Wayne State University, Detroit, MI 48202, USA
}
\author{K. Deng}
\affiliation{Department of Physics, Boston College, 140 Commonwealth Avenue, Chestnut Hill, Massachusetts 02467, USA
}

\author{B. Flebus}
\affiliation{Department of Physics, Boston College, 140 Commonwealth Avenue, Chestnut Hill, Massachusetts 02467, USA
}

\author{J. Sklenar}
\affiliation{Department of Physics and Astronomy, Wayne State University, Detroit, MI 48202, USA
}

\begin{abstract}
In this work we examine synthetic antiferromagnetic structures consisting of two, three, and four antiferromagnetic coupled layers, i.e., bilayers, trilayers, and tetralayers.  We vary the thickness of the ferromagnetic layers across all structures and, using a macrospin formalism, find that the nearest neighbor exchange interaction between layers is consistent across all structures for a given thickness.  Our model and experimental results demonstrate significant differences in how the magnetostatic equilibrium states of even and odd-layered structures evolve as a function of the external field.  Even layered structures continuously evolve from a collinear antiferromagnetic state to a spin canted non-collinear magnetic configuration that is mirror-symmetric about the external field.  In contrast, odd-layered structures begin with a ferrimagnetic ground state; at a critical field, the ferrimagnetic ground state evolves into a non-collinear state with broken symmetry.  Specifically, the magnetic moments found in the odd-layered samples possess stable static equilibrium states that are no longer mirror-symmetric about the external field after a critical field is reached.  Our results reveal the rich behavior of synthetic antiferromagnets.    
\end{abstract}

\maketitle

\section{Introduction}
Synthetic antiferromagnets (SAFs) are a class of magnetic metamaterials comprised of thin ferromagnetic films separated by non-magnetic spacer layers~\cite{duine2018synthetic}.  The spacer layer thickness can be adjusted so that the interlayer coupling, typically originating from the Ruderman–Kittel–Kasuya–Yosida (RKKY) interaction, is antiferromagnetic.  SAFs can be created using thin film deposition techniques, e.g, sputtering, so that various ferromagnetic and interlayer materials can be incorporated into the overall structure.  This experimental flexibility has more recently enabled the incorporation of additional intra/interlayer interactions, such as antisymmetric exchange interactions between layers~\cite{han2019long, fernandez2019symmetry, legrand2020room,yang2019chiral, lonsky2022dynamic}.   

Current interest in SAF materials is being driven from a broad range of research topics in magnetism such as THz generation~\cite{zhang2020terahertz,zhong2020terahertz,wu2021principles}, skyrmions~\cite{legrand2020room,dohi2019formation,chen2020realization}, neuromorphic computing~\cite{siddiqui2019magnetic, yu2020voltage, wang2023spintronic}, and for hosting non-Hermitian phenomena such as exceptional points~\cite{lee2015macroscopic, galda2016parity, liu2019observation, hurst2022non, li2022multitude,jeffrey2021effect,deng2023exceptional}.  Of specific note, magnetization dynamics in SAFs are generating attention due to the accessibility of both optical and acoustic magnons at GHz frequencies~\cite{li2016tunable, wang2018dual, waring2020zero, shiota2022polarization}.  Consequently, SAFs have become an important material platform where  fundamental investigations into magnon-magnon interactions~\cite{sud2020tunable,shiota2020tunable, li2021symmetry, dai2021strong,macneill2019gigahertz} within antiferromagnets can take place.  These studies are partially driving the emerging field of hybrid magnonics~\cite{li2020hybrid,awschalom2021quantum}, where the tunable magnon-magnon interactions that exist in SAFs tuning interactions between magnons and other quasiparticles.

The archetypical SAF structure consists of two magnetic layers separated by a non-magnetic spacer layer.  In this article, we will refer to such structures as bilayers (counting only the magnetic layers in the structure).  Historically, less attention has been paid to the synthesis and characterization of SAFs with more than two coupled antiferromagnetic layers, e.g., trilayers and tetralayers.  In this work, we were motivated to synthesize and characterize SAFs with more than two layers based on our recent computational~\cite{sklenar2021self} and experimental measurements~\cite{subedi2023magnon}, which demonstrated how additional magnon-magnon interactions more readily manifest when the magnons localize in different regions of a SAF structure.  For example, in a tetralayer, pairs of optical and acoustic magnons can tend to localize on surface layers or interior layers of the structure.  In such structures, these magnon-magnon interactions generate characteristic avoided energy level crossings in the magnon energy spectrum.  The frequency where the avoided energy level crossing occurs depends on both the number of layers and the strength of the interlayer exchange interaction.  Thus, a specific goal of this work is to extract the interlayer exchange interaction across a series of structures with various ferromagnetic film thicknesses.

\section{Macrospin Model}
In this work, three magnetic structures are considered.  We will refer to these structures as bilayers, trilayers, and tetralayers.  These structures have, respectively, two, three, and four magnetic layers, each characterized by a nearest neighbor antiferromagnetic interaction.  In all structures, we are able to model the evolution of the net magnetic moment as a function of the external field, $H_0$ with a single parameter, $H_E$.  Here, $H_E$ is the effective exchange field that accounts for the strength of the antiferromagnetic interaction between adjacent ferromagnetic layers.

We consider a SAF stacked in the $z$ direction with the easy-plane $xy$. The external field ($\mathbf{H_0}=H_0\hat{y}$) induced magnetic moment of a given SAF depends on the static equilibrium configuration of a SAF.  For a bilayer, the magnetization of the system can be described using two macrospins denoted by two unit vectors, $\mathbf{\hat m_{A}}$, and $\mathbf{\hat m_{B}}$.  The static equilibrium condition is given by $\mathbf{\hat m_{A(B)}} \times \mathbf{H_{A(B),eff}} = 0$.  Here $\mathbf{H_{A(B),eff}}$ is the effective magnetic field that each macrospin experiences, i.e., $\mathbf{H_{A(B),eff}} = \mathbf{H_0} - H_E\mathbf{\hat m_{B(A)}}$.  In the case of a bilayer, the net effective field magnitude each layer experiences is equivalent due to the symmetry of the system.  As illustrated in Fig. 1 (a), only one equilibrium angle ($\phi_A$) is needed to describe how the magnetization evolves as the external field increases.  The relationship between $\phi_A$, $H_0$, and $H_E$ is given by:
\begin{equation}
    \sin{\phi_A}  = H_0/2H_E.
\end{equation}
In Fig. 1 (a) we plot $\phi_A(H_0)$ assuming that $H_E$ = 500 Oe.

For the tetralayer, four macrospins are needed to describe the magnetic state of the system: $\mathbf{\hat m_{A}}$, $\mathbf{\hat m_{B}}$, $\mathbf{\hat m_{C}}$, and $\mathbf{\hat m_{D}}$. 
The surface layers ($A$ and $D$) experience a different effective field compared to the interior layers ($B$ and $C$) due to the fact that the interior layers are exchange coupled to two layers.  Consequently, two equilibrium angles ($\phi_A$ and $\phi_B$) are used to describe the orientation of the magnetization as $H_0$ increases.  The static equilibrium configuration of the tetralayer is illustrated in Fig. 1 (b).  To obtain the equilibrium angles, we  use the equilibrium condition of the surface layers and the interior layers, i.e., $\mathbf{\hat m_{A(D)}} \times \mathbf{H_{A(D), eff}} = 0$, and $\mathbf{\hat m_{B(C)}} \times \mathbf{H_{B(C), eff}} = 0$.  By considering that the effective fields arise from a combination of $H_0$ and $H_E$, it can be shown that the relationships between $\phi_A$, $\phi_B$, $H_0$, and $H_E$ are given by:
\begin{equation}
    H_0\cos \phi_A - H_E \sin (\phi_A + \phi_B) = 0
\end{equation}
\begin{equation}
    H_0\cos \phi_B - \left[H_E \sin (\phi_A + \phi_B) + \sin(2\phi_B)\right] = 0
\end{equation}
The equilibrium angles can then be obtained by numerically solving Eqs. (2) and (3) for a given value of $H_0$ and $H_E$.  In Fig.  1 (b), we plot $\phi_A(H_0)$ and $\phi_B(H_0)$ for $H_E$ = 500 Oe.
\begin{figure}
    \centering
    \includegraphics[scale=0.28]{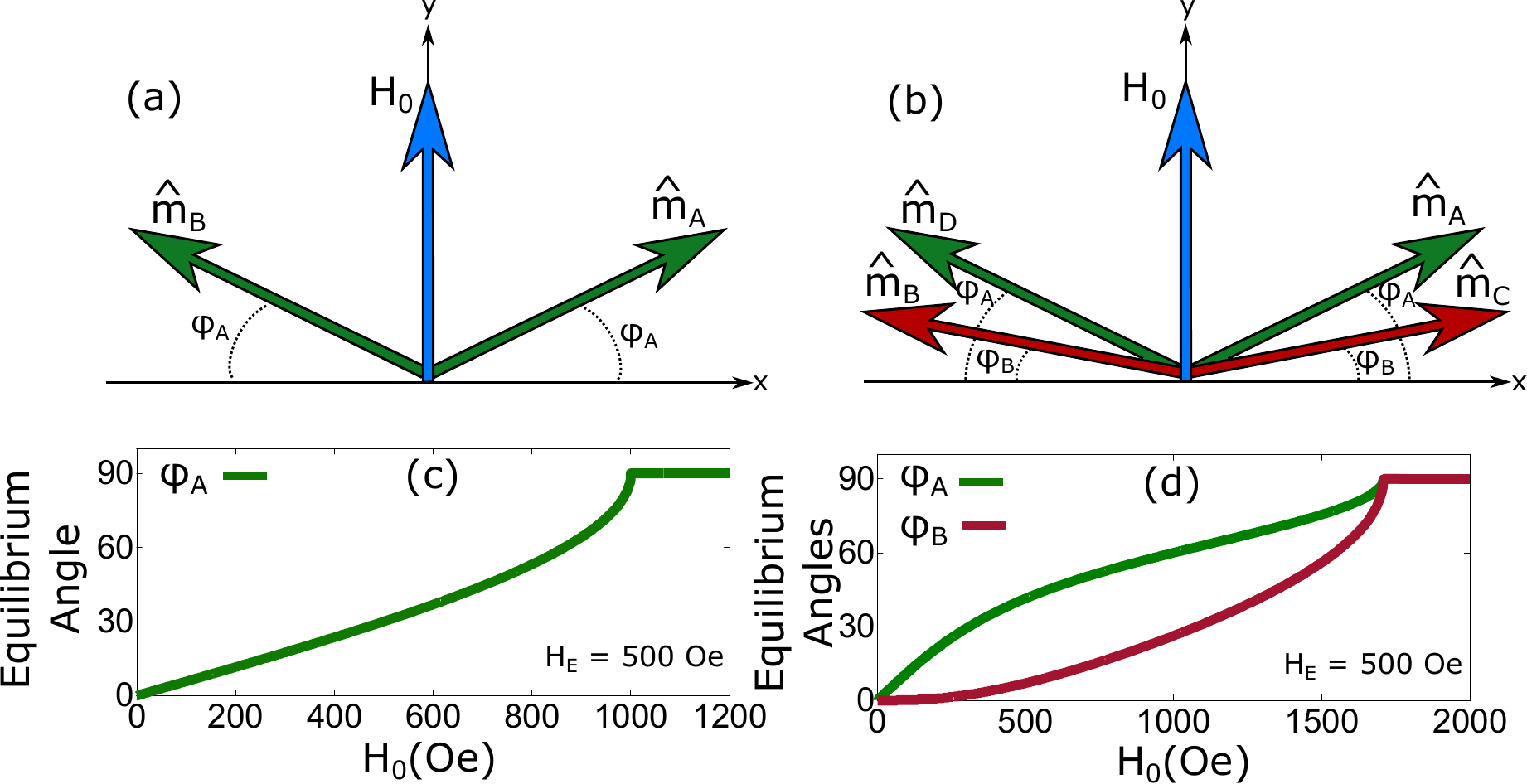}
    \caption{(a) and (b) illustrate the magnetostatic equilibrium configurations for a bilayer and tetralayer in the presence of an external field, $H_0$.  (c) and (d) show the equilibrium angle(s) as a function of external field for the bilayer and tetralayer respectively when $H_E$ = 500 Oe.}
    \label{fig:1}
\end{figure}

 \begin{figure}
    \centering
    \includegraphics[scale=0.35]{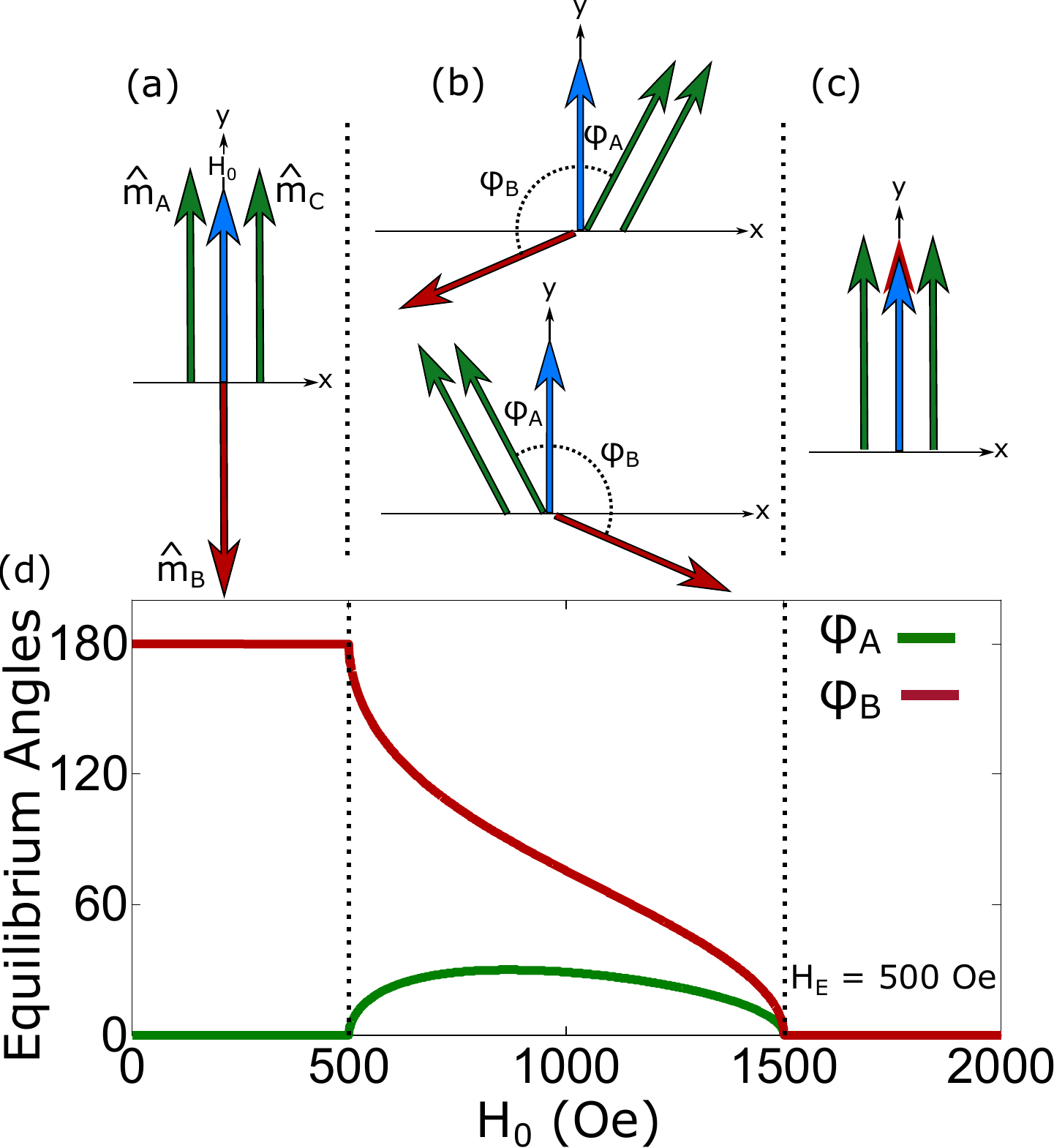}
    \caption{(a), (b), and (c) illustrate three qualitatively different types of static equilibrium configurations that exist within a SAF trilayer.  (a) depicts the low field, ferrimagnetic ground state.  (b) depicts non-collinear symmetry-broken configurations which are stable at intermediate external fields.  (c) depicts the saturated ferromagnetic configuration which occurs at large magnetic fields.  (d) plots the evolution of the equilibrium angles as a function of external field for $H_E$ = 500 Oe.  Note that a discontinuous change in the equilibrium angles at $H_0$ = 500 Oe, when the structure enters the symmetry-broken magnetostatic equilibrium configuration.}
    \label{fig:2}
\end{figure}

The bilayer and tetralayer systems can be thought of as easy-plane antiferromagnetic materials.  The static equilibrium configurations of both bilayers and tetralayers are mirror symmetric about the external field direction for any external field.  With these traits in mind, we now show that odd-layered systems, such as a trilayer, exhibit qualitatively different behaviors compared to even-layered systems.  To describe a trilayer three macrospins are needed: $\mathbf{\hat m_{A}}$, $\mathbf{\hat m_{B}}$, and $\mathbf{\hat m_{C}}$.  The first significant difference between a trilayer and the even-layered systems is that, upon applying and removing an external field, the remnant equilibrium configuration of the trilayer resembles a \textit{ferrimagnetic} ground state.  Specifically, the magnetization of the two surface layers ($A$ and $C$) aligns in the opposite direction of the middle layer ($B$).  This leaves the system with a net magnetic moment which is one-third of the saturated magnetic moment.

\begin{figure*}
    \centering
    \includegraphics[scale=.3]{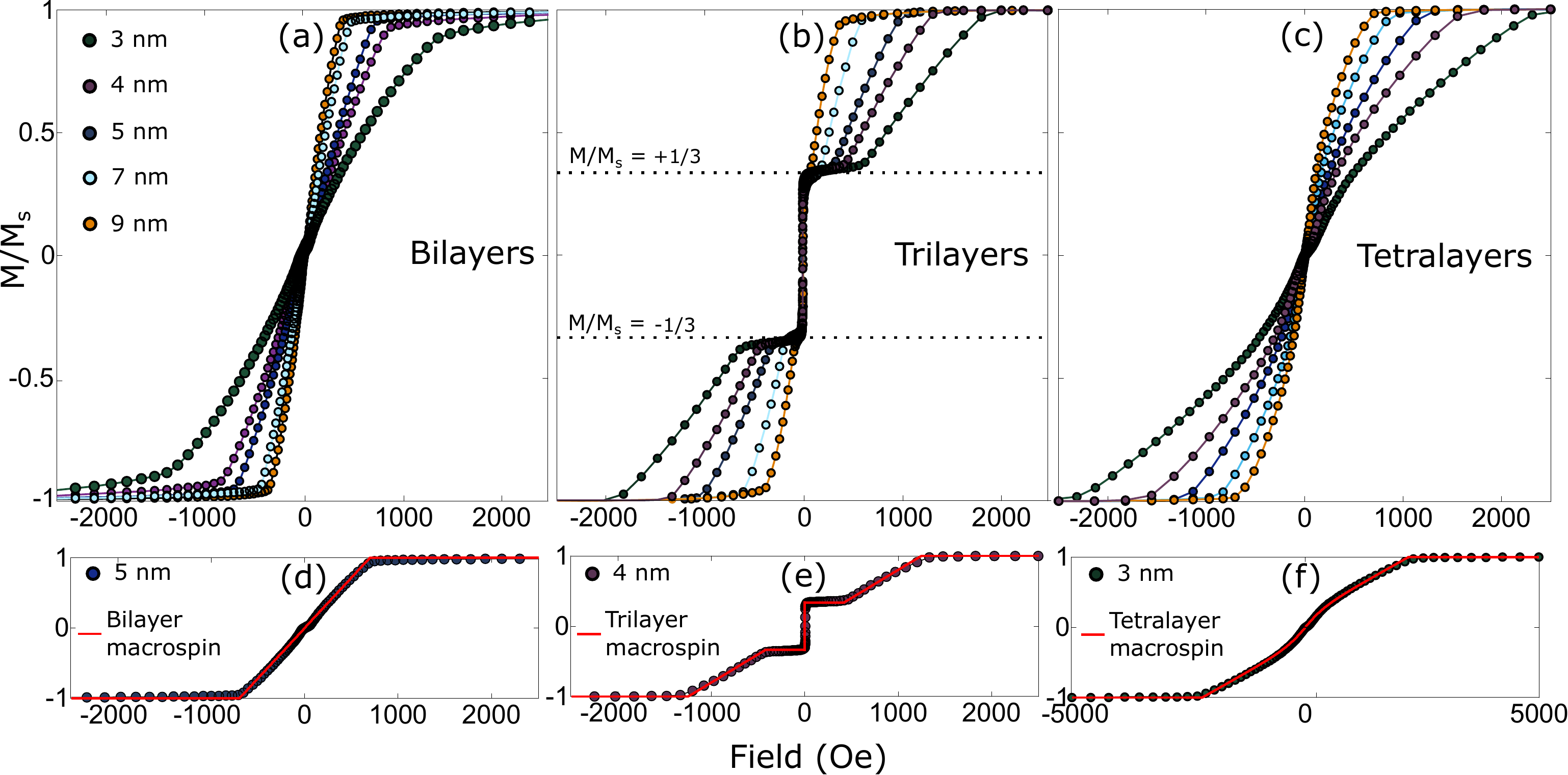}
    \caption{(a), (b), and (c) plot the normalized magnetization as a function of external field for a selection of bilayer, trilayer, and tetralayer samples respectively.  The caption denotes the thickness of the permalloy layers in each of the structures.  (d), (e) and (f) plot the macrospin model prediction on top of the experimental data for a representative bilayer, trilayer, and tetralayer sample where the permalloy was fixed to be 3 nm.  The macrospin model does an excellent job describing the observed behavior and the interlayer exchange field, $H_E$ can be extracted from the the fit between model and experiment.}
\end{figure*}

The ferrimagnetic state found within the trilayer is mirror symmetric about the external magnetic field.  In a trilayer with easy-plane anisotropy, there are only a discrete number of static equilibrium states which are mirror symmetric about the external field.  These states are all collinear to the external field [see Fig. 2 (a) and (c)].  Unlike the bilayer and the tetralayer, the trilayer also possesses stable static equilibrium configurations that are highly non-collinear and are not mirror symmetric about the external magnetic field [see Fig. 2 (b)].  Two equilibrium angles ($\phi_A$ and $\phi_B$) are needed to describe the equilibrium configuration of the trilayer.   By considering the two equilibrium equations $\mathbf{\hat m_{A(C)}} \times \mathbf{H_{A(C), eff}} = 0$ and $\mathbf{\hat m_{B}} \times \mathbf{H_{B, eff}} = 0$, the following pair of equations can be found which are used to define symmetry-broken stable magnetization configurations:
\begin{equation}
H_0 - H_E\left[\cos{\phi_B} + \cot{\phi_A}\sin{\phi_B}\right] = 0,
\end{equation}
\begin{equation}
H_0 - 2H_E\left[\cos{\phi_A} + \cot{\phi_B}\sin{\phi_A}\right] = 0.
\end{equation}

In Fig. 2 (d) we show the stable equilibrium angles for a trilayer, assuming that $H_E$ = 500 Oe.  One can clearly see that there are three distinct regions that describe the trilayer.  Initially, when $H_0$ $<$ $H_E$, the stable static equilibrium configuration is the collinear ferrimagnetic  state.  For $H_E < H_0 < 3H_E$ the collinear states become unstable, and the trilayer spontaneously evolves into one of the two degenerate symmetry-broken equilbrium states.  In these symmetry broken states, the magnetization of the surface layers will initially, spontaneously, rotate clockwise (or counterclockwise) away from the external magnetic field.  Simultaneously, the middle layer will rotate towards the external field in the same direction that the surface layers have spontaneously rotated away from the field as shown in Fig. (b).  The evolution of the ground state magnetization as a function of the field $H_0$ is a classical example of \textit{spontaneous} symmetry breaking.  At $H_0$ $=$ $H_E$ the single energy minimum bifurcate into two new minima, corresponding to configurations that can be mapped to each other via mirror reflection $m^x_i\rightarrow -m^x_i$, as shown in Fig. 2 (b).  When $H_0$ $>$ $3H_E$, the two free energy minima merge into one corresponding to a mirror symmetric configuration, i.e., a ferromagnetic collinear ground state.

\section{Experimental Results}

All the samples were deposited using DC magnetron sputtering with Ar plasma onto natively oxidized silicon substrates at room temperature. Sputtering is performed in a high vacuum system with a base pressure near $3\times10^{-9}$ Torr. Deposition occurs in the presence of a 3 mTorr atmosphere of Ar gas. We synthesized Pt/(Py/Ru)$_x$/Py/Pt structures where $x$ denotes the number of (Py/Ru) repetitions.  For all samples, the thickness of Ru was kept constant at 1 nm where we find the interaction between permalloy layers to be antiferromagnetic. The thickness of  permalloy was varied between 2-13 nm.  The top and bottom layers of the overall structure are fixed to be 6 nm of Pt. We found that the residual ferromagnetic moment created due to uncompensated magnetization was suppressed when the overall structure contained a top and bottom Pt layer. The DC power used in the growth of permalloy and ruthenium layers is 50 W, and the DC power is 30 W for platinum.

To understand the magnetic behavior of our samples, we used a magnetic properties measurement system (MPMS) to measure the magnetic moment of a given sample as a function of magnetic field at room temperature.  Every sample that was measured had a diamagnetic background signal due to the sample substrate and sample holder.  We subtracted the diamagnetic component out of every measurement by measuring just the diamagnetic signal over an external field range where the magnetization of the sample was clearly saturated.  In the case of our samples, this field range was 20 kOe - 40 kOe. In Fig. 3 (a), (b), and (c) we plot the normalized magnetization as a function of external field for a representative set of bilayer, trilayer, and tetralayer samples, respectively.  For all samples, we find that as the thickness of the permalloy decreases, a stronger external magnetic field is needed to saturate the magnetization.  The bilayer and tetralayer samples do not exhibit any pronounced remnant magnetization.

\begin{figure}
    \centering
    \includegraphics[scale=.2]{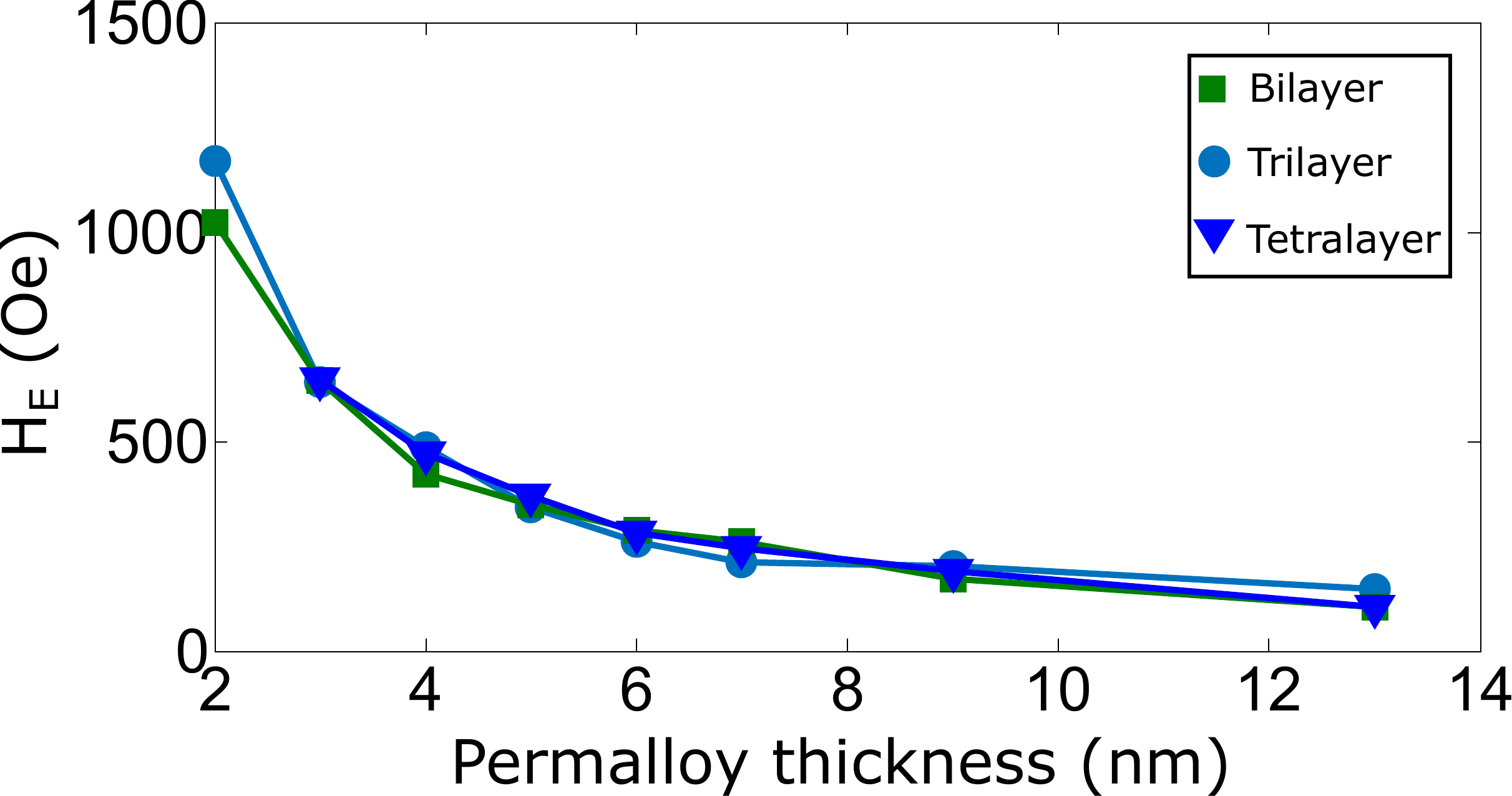}
    \caption{The values of the interlayer exchange field, $H_E$ are summarized as a function of the permalloy thickness for all of the bilayer, trilayer, and tetralayer samples that were considered in this work.}
\end{figure}

The magnetization of the bilayer samples increases linearly as a function of the magnetic field.  Fig. 3 (d) shows a representative comparison of the macrospin model with the experimental data for a permalloy thickness of 3 nm.  For the case of the bilayer, $M/M_s = \sin\phi_A$.  Thus, using eq. (1), the slope of linear curve describing the magnetization as a function of field is equal to $1/2H_E$.  The magnetization curves of the tetralayer increase monotonically in a non-linear manner. The magnetization first increases more rapidly at low fields before a ``knee'' like feature is observed, followed by a slower increase of the magnetization as it approaches saturation.  This behavior is captured completely by the macrospin model, and a representative comparison of the macrospin model to the experimental data for a sample with permalloy layers 3 nm thick is shown in Fig. 3 (f). We emphasize that the field-dependent evolution of the magnetization for both the bilayer and tetralayer are described by the non-collinear magnetostatic equilibrium configurations that are symmetric about the external magnetic field as shown in Fig. 1.

The trilayer results qualitatively contrast  the bilayer and tetralayer results in several notable ways.  First, all samples have a remnant magnetic moment that is near $\pm 1/3$ the saturated magnetic moment.  As the external field is swept from a negative to positive value (or vice versa), the moment of a trilayer will suddenly switch between these two states.  As the external field increases to a higher magnitude, the magnetization initially does not increase/deviate very much beyond the remnant value.  This behavior leads to the appearance of two ``plateaus'' in the magnetization curves near  $\pm 1/3$.  Eventually, a large enough external field will cause the magnetization to increase abruptly in a linear manner.  Qualitatively, in all trilayer samples we see a linear increase in magnetization from $M/M_s = 1/3$ to $M/M_s = 1$ after a large enough external field is reached.

As shown in Fig. 3 (e), the macrospin formalism does an excellent job of describing the behavior of the trilayers.  The plateau regions, where $M/M_s$ is near $\pm1/3$, are consistent with the low-field ferrimagnetic configurations calculated in Fig. 2.  The regions of the magnetization curves, where the magnetization increases linearly, are very well described by the symmetry-broken states.  In fact, in the absence of the symmetry broken static equilibrium states, one would not expect a continuous variation in the magnetization from $M/M_s = \pm1/3$ to $M/M_s = \pm1$.  If the symmetric magnetization states were the only valid static equilibrium configurations, one would expect a sudden increase in the magnetization from $M/M_s = \pm1/3$ to $M/M_s = \pm1$ once the strength of the external field forced the ferrimagnetic configuration to be an unstable equilibrium relative to the full saturated configuration. 

A summary of the interlayer exchange fields, $H_E$, across all structures and all thicknesses is found in Fig. 4.  In general, we find that the extracted values of $H_E$ for a given thickness of permalloy are consistent across all thicknesses.  However, we uncovered two notable situations that may suggest limitations of the macrospin model.  First, we found that the macrospin model does not adequately describe the magnetization curves of a 2 nm tetralayer, especially as the magnetization became nearly saturated.  Hence, no reliable data point for this structure is shown in Fig. 4.  Additionally, we found that when trying to extract $H_E$ from the trilayer using the entire field range, we would consistently obtain values of $H_E$ that were 20 - 80 Oe smaller than what was extracted from the bilayer and tetralayer.  However, by restricting our analysis of the trilayer data set to the region where the magnetization is linearly increasing, $H_E < H_0 < 3H_E$, we extracted values that were consistent with what was extracted from the bilayer and tetralayer analysis.  A closer examination of the trilayer samples hints at behavior not fully captured by the macrospin model, which may be responsible for this deviation.  In particular, in the low-field ferrimagnetic region, the experimental values of $M/M_s$ tend to slowly increase from 0.33 to 0.37 instead of staying pinned on a value of 0.33.  Investigating the origin of non-macrospin related effects in odd-layered structures would thus appear to be an interesting topic for future studies.

\section{Summary and Outlook}

In this work we have successfully synthesized, characterized, and modeled the magnetostatic properties of SAF bilayers, trilayers, and tetralayers based on permalloy and ruthenium.  The macrospin model used to describe our samples is successfully applied over a range of permalloy thicknesses between 2-13 nm.  We find that the interlayer exchange interaction is enhanced and greater than 1 kOe for the SAF structures with permalloy thickness of 2 nm.  By measuring the magnetization as a function of the external field for trilayers, we observe a behavior described by the magnetization having highly non-collinear configurations with a broken mirror symmetry about the external magnetic field.  Moving forward, investigations into the magnon energy spectrum of the trilayer structures will be a promising direction to pursue.  Already, we have found that new magnon-magnon interactions, mediated by spin-pumping, significantly alter the energy spectrum within tetralayers (as compared to bilayers)\cite{subedi2023magnon}.  With this newly established understanding of non-collinear configurations in trilayers, we are poised for similar studies to take place.

\section{Acknowledgements}

Experimental work at Wayne State University was supported by  U.S.  National  Science  Foundation   under  award  DMR-2117487.  M.M.S and J.S. acknowledge support from the National Science Foundation under DMR-2328787.  B. F. acknowledges support from the National Science Foundation under DMR-2144086.

\end{document}